\documentstyle[12pt]{article}
\newcommand{\be}[3]{\begin{equation}  \label{#1#2#3}}
\newcommand{\ee}{ \end{equation}}
\newcommand{\ba}{\begin{array}}
\newcommand{\ea}{\end{array}}

\begin{document}


\thispagestyle{empty}
\rightline{UG-5/96}
\rightline{HUB-EP-96/15}
\rightline{hep-th/9605216}
\rightline{May 1996}
\vspace{1truecm}
\centerline{\bf A Note on Intersecting $D$-branes and Black Hole Entropy}
\vspace{1.2truecm}
\centerline{\bf Klaus Behrndt}
\vspace{.5truecm}
\centerline{Humboldt-Universit\"at, Institut f\"ur Physik}
\centerline{Invalidenstra\ss e 110, 10115 Berlin}
\centerline{Germany}
\vspace{.3truecm}
\centerline{and}
\vspace{.3truecm}
\centerline{{\bf Eric Bergshoeff}}
\vspace{.5truecm}
\centerline{Institute for Theoretical Physics}
\centerline{Nijenborgh 4, 9747 AG Groningen}
\centerline{The Netherlands}
\vspace{1.2truecm}


\vspace{.5truecm}

\begin{abstract}
\noindent
In four dimensions there are 4 different types of extremal Maxwell/scalar
black holes characterized by a scalar coupling parameter $a$ with 
$a=0,1/\sqrt{3} , 1 , \sqrt{3}$. These black holes can be
described  as intersections of ten--dimensional 
non-singular Ramond-Ramond objects,
i.e.~$D$-branes, waves and Taub-NUT solitons. Using this description 
it can be shown that the four--dimensional
black holes decompactify near the core to higher--dimensional {\em
non-singular} solutions. In terms of these higher--dimensional
non-singular solutions
we define a non-vanishing entropy for all four 
black hole types
from a four--dimensional point of view.
\end{abstract}

\newpage
\noindent {\bf 1.\ Introduction}
\bigskip

\noindent
A common way to classify four--dimensional
Maxwell/scalar black hole (BH) solutions is to
specify the coupling of the scalar fields to the gauge fields.
In the simplest case of only one
scalar field and one gauge field this coupling is characterized by
a single parameter $a$ and the action in the Einstein frame is
given by
\be010
 S_{4d} = \frac{1}{16 \pi G_4} \int d^4 x \sqrt{|g|} \{-R + 
 2 (\partial \phi)^2  + e^{-2 a \phi} F^2 \} \ ,
\ee
where $G_4$ is the 4--dimensional Newton constant.
There exists four different types of extremal\footnote{In this letter 
we consider only extremal solutions.} black hole solutions, which are 
defined in terms
of a function $H(\vec{x})$ which is harmonic on the 3--dimensional
transverse space. The metric of these solutions is given 
by
\be020
\ba{llll}
a=0 & : & ds^2 = H^{-2} dt^2 - H^2 d\vec{x}^2 \quad , & e^{-2 \phi} =1 \ , \\
a=1/\sqrt{3} & : & ds^2 = H^{-3/2} dt^2 - H^{3/2} d\vec{x}^2 \quad , 
    & e^{\pm 2 \phi /\sqrt{3} } =\sqrt{H}  \ , \\
a=1 & : & ds^2 = H^{-1} dt^2 - H d\vec{x}^2 \quad , & e^{\pm 2 \phi} =H  \ ,\\
a=\sqrt{3} & : & ds^2 = H^{-1/2} dt^2 - H^{1/2} d\vec{x}^2 \quad , 
 & e^{\pm 2 \phi/\sqrt{3}} = \sqrt{H}  \ .
\ea
\ee
\noindent The harmonic function $H(\vec{x})$ is given by
\be222
H(\vec{x}) = 1 + \frac{r_h}{r}\, ,
\ee
where $r^2 = \vec{x} \cdot \vec{x}$    
and $r_h$ is proportional to the charge. 
These solutions have been generalized to different harmonic functions
in \cite{ra} (for $a=0$ this generalization has been given in \cite{cv/ts}). 
For a recent discussion of these solutions as 
bound states, see \cite{du/ra}.
The four solutions (\ref{020}) are also known as 
\medskip

\begin{tabular}{lll}
$a=0$ & :  & 4d Reissner-Nordstom (RN) solution, \\
$a=1/\sqrt{3}$ & : & 5d RN  ($\phi$ is a modulus field), \\
$a=1$ & :  & dilaton black hole ($\phi$ has standard dilaton coupling), \\
$a=\sqrt{3}$ & : & 5d KK black hole. 
\end{tabular}   \smallskip
\medskip

For $a\ne 0$ the gauge fields can be electric or magnetic. The two
possibilities correspond to different signs of the scalar field
$\phi$. In formula (\ref{020}) the ``+'' sign corresponds to the
magnetic case. On the other hand, the $a=0$ RN solution is dyonic.
It turns out that in four dimensions only the $a=0$ RN solution is non-singular
at the horizon $r=0$. However, following \cite{gi/ho/to},
also the $a\ne 0$ solutions can be understood
in a non-singular way, in the sense that they follow from the
dimensional reduction of the following higher--dimensional
non-singular solutions: 
\medskip

\begin{tabular}{lll}  
$a=1/\sqrt{3}$ & : & 5d RN electric black hole or magnetic string\, , \\
$a=1$ & : & 6d self-dual string\, , \\
$a=\sqrt{3}$ & : & 10d self-dual $D$-3-brane\, . 
\end{tabular}   \smallskip
\medskip

All solutions can be understood as intersections of branes in 10 or 11
dimensions \cite{pa/to,ts,ba/la,be/be/ja,ga/ka/tr,kh/lu/po}. Most of
them are singular.  But there are some objects which are
non-singular. In ten dimensions these are: $D$-3-branes, gravitational
waves and Taub-NUT solitons\footnote{We do not discuss the 2- and
5-brane solutions of the 11--dimensional $M$-theory, which are also
non-singular. They appear in 10 dimensions as a subset of the
$D$-branes which we prefer to consider. There are also other branes
that have a non-singular metric , e.g.~the $p=-1$ brane which is a
wormhole in the string frame \cite{gi/gr/pe}. However, these solutions
are not asymptotically flat and/or have a singular dilaton.  We will
ignore these solutions as well.}. Only the first object carries
Ramond--Ramond (RR) charges and is part of the $D$-branes. The wave
and Taub-NUT solitons are neutral and do not fit in the standard brane
picture.  They appear as the $T$-dual of the fundamental NS string and
NS 5-brane. Since they are solutions of pure gravity (without scalars
and gauge fields) we can regard them as solutions of the RR
sector. Although they are neutral in 10 dimensions their metric is
non-diagonal and yields KK gauge fields in lower dimensions.  The aim
of this letter is to understand the black holes (\ref{020}) and their
higher--dimensional origins in terms of an intersection of these
ten--dimensional non-singular objects. In analogy to \cite{ra,du/ra}
we can see this orthogonal intersection as bound states of fundamental
$a=\sqrt{3}$ states (a single object or brane).

\bigskip

\noindent{\bf 2.\ Black Holes as intersecting $D$--branes}
\bigskip

\mbox{Since} the metric of a single $D$--brane solution in ten dimension
involves the square root of a harmonic function and the power of the
harmonics in (\ref{020}) is at most two we know that in order to
describe the BH's in (\ref{020}) as intersections we need at most 4
objects in 10 dimensions, each defined by its own harmonic
function. At this point one could ask whether there exist more
BH's, e.g.~described by 5,6,.. harmonic functions. 
It is easy to see that this is not possible. 
An odd number of harmonic functions is ruled out by our restriction
to find a non-singular intersection, we need an even number to keep
the compactification radii finite. But why not 6 harmonic functions?
A further restriction is that for 2 non-trivial
functions only the self-dual string describes a non-singular
intersection, i.e.~any pair of harmonics has to describe this
string. For 4 objects we can build 6 pairs and the corresponding
strings fit nicely in the 6 internal directions. On the other side
for 6 objects we can build 15 pairs or strings and there is no
way to put these strings into the internal space. Some of them
have to lie in the same internal direction resulting in a singular scalar
field. With similar arguments one can discard also the higher cases.
Thus let us come back to the case of 4 intersecting objects.
For the $a=0$ case all functions are non-trivial. Since this
case is dyonic there are two possible intersections: ${\it (3 \times 3
\times 3 \times 3)}$ where all objects are dyonic or ${\it (3 \times 3
\times \tilde{1} \times \tilde{5})}$ where the electric (KK) charge
coming from the wave ($= {\it \tilde{1}}$) compensates the magnetic
charge from the Taub-NUT soliton ($= {\it \tilde{5}}$)\footnote{We do
not consider the case ${\it (\tilde{1} \times
\tilde{5} \times \tilde{1} \times \tilde{5})}$ where none of the
components exhibits a horizon.  We denote the wave with ${\it
\tilde{1}}$ to indicate that it is T-dual to the fundamental string
${\it 1}$ and the Taub-NUT soliton with ${\it \tilde{5}}$ since it is
T-dual to the solitonic 5-brane ${\it 5}$.}.  We consider these two
cases separately.

\bigskip

\noindent
{\bf (i)\ the ${\it (3 \times 3 \times 3 \times 3)}$ case} \bigskip

\noindent
We take  4 intersecting $D$-3-branes with metric given by
\cite{ts,ba/la}\bigskip
\be030
\ba{lcl}
ds^2 &=&\frac{1}{\sqrt{H_1 H_2 H_3 H_4}} dt^2  - 
       \sqrt{H_1 H_2 H_3 H_4}d\vec{x}^2 
   - \sqrt{\frac{H_1 H_2}{H_3 H_4}} dx_4^2  
   - \sqrt{\frac{H_1 H_3}{H_2 H_4}} dx_5^2 - \\ 
  &&  - \sqrt{\frac{H_1 H_4}{H_2 H_3}} dx_6^2 - 
   \sqrt{\frac{H_2 H_3}{H_1 H_4}} dx_7^2 -
    \sqrt{\frac{H_2 H_4}{H_1 H_3}} dx_8^2 -
   \sqrt{\frac{H_3 H_4}{H_1 H_2}} dx_9^2\, .
\ea
\ee
\medskip

\noindent The electric gauge field components are
\be040
\ba{rl}
F \sim & dt \wedge \left( d H_1^{-1} \wedge dx_7 \wedge dx_8 \wedge dx_9
  +  dH_2^{-1} \wedge dx_5  \wedge dx_6 \wedge dx_9 + \right. \\
  & + \left.  dH_3^{-1} \wedge dx_4 \wedge dx_6 \wedge dx_8 
  +  dH_4^{-1} \wedge dx_4  \wedge dx_5 \wedge dx_7
  \right) \ .
 \ea
\ee
The magnetic components can be obtained by using the self-duality
condition in $D=10$. The special cases (\ref{020}) appear in the limit
\be050
\ba{rcl}
a=0 & : & H_1 = H_2 = H_3 = H_4 =H \ , \\
a=1/\sqrt{3} & : & H_1 = H_2 = H_3 = H \quad , \quad H_4 =1 \ ,  \\
a=1 & : & H_1 = H_2 = H \quad , \quad H_3 = H_4 = 1 \ , \\
a=\sqrt{3} & : & H_1 = H \quad , \quad H_2 = H_3 = H_4 = 1  \ .
\ea
\ee
The harmonic function $H$ depends on all overall transversal
coordinates, i.e.\ the coordinates for which the metric has no $H$ in
the denominator.  The last case, e.g.\ , is the single 3-brane with a
harmonic function depending on: $\vec{x}, x_4, x_5, x_6$ and for the
$a=1$ case $H$ is harmonic with respect to the coordinates: $\vec{x},
x_4$. The solution (\ref{030}) is non-singular for the case
$a=0,1,\sqrt{3}$. However, the case $a=1/\sqrt{3}$ is singular.
We next consider the second intersection.

\bigskip

\noindent
{\bf (ii)\ the ${\it (3 \times 3 \times \tilde{1} \times \tilde{5})}$ case} 
\bigskip

\noindent
We intersect two  $D$-3-branes with a wave and a Taub-NUT
soliton. The resulting metric is given by
\be060
\ba{rcl}
ds^2 & =& \frac{1}{\sqrt{H_1 H_2}} du ( dv - \tilde{H}_1 du ) - 
   \sqrt{H_1 H_2}\left[ \frac{1}{\tilde{H}_5} (dx_5 + \vec{V} d\vec{x})^2 +  
   \tilde{H}_5 d\vec{x}^2\right] - \\
&&  - \sqrt{\frac{H_1}{H_2}} \left( dx_6^2 + dx_7^2 \right) 
  - \sqrt{\frac{H_2}{H_1}} \left( dx_8^2 + dx_9^2 \right)\, .
\ea
\ee
The electric part of the field strength is given by
\be070
F \sim  dv \wedge du \wedge \left( d H_1^{-1} \wedge dx_8 \wedge dx_9
  +  dH_2^{-1} \wedge dx_6 \wedge dx_7 \right)\, ,
\ee
where $v = t + x_4$, $u = t - x_4$ and $\vec{\nabla} \tilde{H}_5 =
\vec{\nabla} \times \vec{V}$. This is a non-diagonal intersection
yielding in 4 dimensions two $D$-brane or RR charges and two KK
charges.  Compactifying this model to 6 dimensions and performing a
type IIB $SL(2,R)$ transformation this model coincides with the
solution discussed in \cite{cv/ts}, which is self-dual under
$T$--duality as well as string/string duality.  In 10 dimensions the
wave lying on the common world volume of the $3$-branes and the
Taub-NUT soliton in the overall transversal space correspond to
additional momentum modes in the internal space. Again we get the 4
BH's (\ref{020}) if we choose the harmonic functions properly:
\be080
\ba{rcl}
a=0 & : & H_1 = H_2 = \tilde{H}_1 = \tilde{H}_5 =H  \ ,\\ 
a=1/\sqrt{3} & : & H_1= H_2 = \tilde{H}_1 = H \quad , 
   \quad \tilde{H}_5 =1 \qquad (\mbox{electric}) , \\ 
 &&   H_1= H_2 = \tilde{H}_5 = H \quad , 
   \quad \tilde{H}_1 =1 \qquad (\mbox{magnetic}) , \\
a=1 & : & H_1 = H_2 = H \quad , \quad \tilde{H}_1 = \tilde{H}_5 = 1 \ , \\
a=\sqrt{3} & : & H_1 = H \quad , \quad \tilde{H}_1 = \tilde{H}_5 = H_2 = 1 \ .
\ea
\ee
As before, we assume that the harmonic functions depend only on the overall
transversal coordinates. The last two cases in (\ref{080}) 
describe the same intersections as
before. In contrast to the intersection of 4 $D$-3-branes, all cases, 
including the
$a=1/\sqrt{3}$ case, are non-singular if we approach
the horizon at $r=0$. 
\bigskip

\noindent {\bf 3.\ Entropy}
\bigskip

Our aim is to use the higher--dimensional interpretation in
terms of intersecting $D$--branes to discuss the entropy of the
four--dimensional black holes. To explain the main idea we first consider
the electric $a=1/\sqrt{3}$ solution in more 
detail. In this case after a trivial reduction (yielding no KK scalars)
the 5d solution is
\be090
ds^2 = \frac{1}{H^2} dt^2 - H (dx_5^2 + d\vec{x}^2) \quad , \quad 
F_{0m} \sim \partial_m H^{-1}\, ,
\ee
where $x_m = (x_5,\vec{x})$ and
\be333
H(x_5, \vec{x}) = 1 + \frac{r_h^2}{\rho^2}\, , 
\ee
with $\rho^2 = x_5^2 +
r^2$. This is the electric 5d RN solution.
The standard $a=1/\sqrt{3}$ BH is obtained after compactification over
$x_5$. Hence, we have to assume that $H$ is periodic: $x_5 \sim x_5 + 
2 \pi R$. Then we can make the standard ansatz for $H$ as a periodic array 
\cite{ga/ha/li}
\be100
H = 1 + \sum_{n= - \infty}^{+\infty} \frac{r_h^2}{r^2 +
   (x_5 + 2 \pi n R)^2} = 1 + \frac{r_h^2}{2R \, r}
   + {\cal O}(e^{-\frac{r}{R}}) \ .
\ee
Thus, away from the origin the dependence on $x_5$ is exponentially
suppressed, this direction is compactified on a circle with radius
$R$.  On the other side near the horizon ($\rho=0$) one can ``feel''
the $x_5$ dependence and the solution decompactifies to its 5d
origin. The philosophy is the same as for the example of rotating BH's
discussed in \cite{ho/se}.  The asymptotic behaviour of the metric
near the horizon is given by
\be110
ds^2 \rightarrow \left( \frac{\rho^2}{r_h^2} \right)^2 \, dt^2 - 
r_h^2 \left( \frac{d\rho}{\rho} \right)^2 - 
 r_h^2 d\Omega_3^2 = e^{4 \eta / r_h} dt^2 - d\eta^2 - r_h^2 d\Omega_3^2\, ,
\ee
where $\rho/r_h = e^{\eta / r_h}$ and $r_h$ is the radius of the $S_3$ sphere.
Hence, the asymptotic geometry is: $(\mbox{de Sitter})_2 \times S_3$ which
is non-singular. In analogy to this procedure one finds for the
other cases \cite{gi/ho/to}:
\be111
\ba{rcl}
a=0 &:& (\mbox{AdS})_2 \times S_2 \ ,\\
a=1/\sqrt{3} &:& (\mbox{AdS})_2 \times S_3 \qquad (\mbox{5d\ 
                                  electric\ RN\ BH}), \\
&& (\mbox{AdS})_3 \times S_2  \qquad (\mbox{5d\ magnetic\ RN\ string}) , \\
a=1 &:& (\mbox{AdS})_3 \times S_3 \qquad (\mbox{6d\ self-dual\ string}) , \\
a=\sqrt{3} &:& (\mbox{AdS})_5 \times S_5  \qquad (\mbox{10d\ self-dual\ 
                                       3-brane})\, ,
\ea
\ee
where $(\mbox{AdS})_n$ is the $n$-dimensional anti de Sitter space.
These asymptotic limits arise also in the extreme limit of
non-extremal black $p$-branes \cite{ho/st}.  In \cite{gi/ho/to} it has
been shown that it is possible to extend the solutions through the
horizon. As a result it was found that for the 4d ($a=0$) and the 5d
electric RN BH there is an interior region with a curvature
singularity. The other cases, however, are completely singularity
free, i.e.~they describe a space time without any singularities but
with a horizon.

\bigskip

Our purpose is to use the description of the BH's given in (\ref{020})
as intersections to define an entropy for
all BH's in (\ref{020}). For every black hole solution one defines the
entropy via the Bekenstein-Hawking formula
\be120
S = \frac{A}{4} \ ,
\ee
where $A$ is the area of the horizon ($r=0$) and we set the 4d Newton 
constant $G_4=1$. Applying this formula naively to the solutions in
(\ref{020}), i.e.\ without allowing a dependence of the
harmonic function on the internal coordinates via
a periodic array, we find
\be130
A = \left( \int_{S_2} R^2(r) d\Omega \right)_{r=0} =
\left( \int_{S_2} \sqrt{H^{\frac{2}{1+a^2}} r^2} d\Omega \right)_{r=0}
 \rightarrow \left( \omega_2 \, r_h^{\frac{2}{1+a^2}} \, 
 r^{\frac{a^2}{1+a^2}} \right)_{r=0}\, ,
\ee
where $R(r)$ is the radius of the sphere for fixed value of $r$ and
$\omega_2=4\pi $ is the 2d unit-sphere volume. Thus, one gets only for
the RN solution ($a=0$) a non-vanishing Bekenstein-Hawking entropy:
$S=2\pi Q^2$ for $r_h = \sqrt{2} \, Q$. Since on the other side the
entropy has the statistical interpretation of counting all possible
states this is not the desired result. As discussed e.g.~in
\cite{du/ra}, it seems strange, that only the case of 4 non-vanishing
charges yields a non-vanishing entropy, whereas for all other cases
one gets zero. However, one can argue that this shortcoming is simply a
consequence of an ill--defined perturbation theory. Either the theory is
in the strong coupling regime (for the magnetic solutions) indicating
a failure of the string perturbation theory (large string coupling
constant $g_s$) or there is a curvature
singularity in the string metric (electric solutions) indicating a
break down of the low energy limit (large $\alpha'$ corrections). Both
perturbation series are under control for the RN solution after
generalizing it to an independent electric ($Q$) and a magnetic ($P$)
charge, which is equivalent to two independent harmonic
functions. Then, on the horizon we have for the string coupling
constant $g_s^2 = e^{2 \phi} \sim (\frac{P}{Q})$ and the string
metric at the horizon is regularized by the magnetic charge,
which effectively corresponds to a renormalization of $\alpha'$
\cite{cv/ts} $\alpha^\prime \rightarrow \frac{\alpha^{\prime 2}}{P^2}$. 
So, assuming that $Q \gg P \gg 1$, i.e.~$\alpha^\prime, g_s \ll 1$
the theory is well defined even on the
horizon.  Here, we argue that similar to the RN case one can keep also
for the other cases both perturbation expansions under control by
allowing the maximal possible dependence of the harmonic functions
on the transversal
coordinates and assuming that the charges are large. As explained for
the electric $a=1/\sqrt{3}$ case this has the consequence that the
solution decompactifies to its higher-dimensional origin.  The
asymptotic geometry is given by (\ref{111}) and as in 4 dimensions the
entropy is defined by integrating over the spherical part. However,
this gives not the total entropy, but the entropy per unit world
volume. Note that the $AdS$ part in (\ref{111}) consists of the
radius and world volume, which are kept fixed in this calculation
(no integration over these coordinates).
To consider this quantity is also motivated by the suggestion
of \cite{fe/ka} that the entropy is given by the
minima of the susy central charge, which in turn is equal to the mass
per unit world volume (Bogomol'nyi bound).  For the total entropy we
have to keep in mind that on the one hand the decompactified branes are
infinitely extended objects and on the other hand the world volume
components of the target space metric vanish on the horizon, i.e.\ one
has to define a suitable limit.  Keeping the brane compactified,
i.e.~wrapped around a torus results in a vanishing total entropy.
One can get a non-trivial result if one goes to the non-extremal case.
For the $D$-3-brane this has been investigated in \cite{gu/kl/pe}.

Let us now compare the different entropy contributions. First, we note
that the radius of all horizons is given by $r_h$, which can be
expressed by the electric and magnetic charges. Since the 4 dimensional
solution can be electric as well as magnetic, let us distinguish  
between both charges. Then for $a=0, 1, \sqrt{3}$, after integrating
over the different spherical parts $S_k$ (i.e.~$k = 2+a^2$) the entropy 
per unit world volume can be written as
\be140
\ba{rl}
& S = \frac{A}{4 \, G_k} = \frac{1}{4 \, G_k} \int_{S_k}
(r_h)^k d\Omega_k = \pi (r_h)^k \\[2mm]
\mbox{with:}& (r_h)^2 = 
   4  \sqrt{\left(( \vec{n} + \frac{1}{2} \vec{Q}) L (\vec{n} + 
   \frac{1}{2} \vec{Q})\right) \left( (\vec{p} + \frac{1}{2} \vec{P}) 
   L (\vec{p} + \frac{1}{2} \vec{P})\right)} \ .
\ea
\ee
where $L$ is the metric in the $O(d,d)$ space\footnote{In this
formula we have already included the different ways of embedding the
intersection into the 10d space, which causes this $O(d,d)$
structure.}  and $\vec{n}$ and $\vec{p}$ are arbitrary unit vectors 
($\vec{n} L \vec{n} = \vec{p} L \vec{p} =1$).  With
$G_k$ we take into account that the Newton constant has to be
rescaled when one compares expressions in different dimensions (see
e.g.~\cite{ho}). In our normalization in 4 dimensions we have $G_4=1$.

\mbox{From} the black holes in (\ref{020}) we do not see that some
of these solutions have a hidden part of the horizon. Instead, to
measure the area of the horizon one could draw an $S_2$ around the
origin and shrink the radius untill we reach the horizon. 
By this procedure one would define a common $S_2$ cut through
all horizons. Integrating over this $S_2$ only (i.e.~$k=2$ in
(\ref{140})) and assuming that $\vec{n} \cdot \vec{Q}=
\vec{p} \cdot \vec{P}=0$ yields the entropy formula suggested by 
\cite{la/wi}. In this reference one can find an
explanation of the case where we have no charges at all. In this case,
the term $S_c(Q=P=0) = 2 \sqrt{\frac{\pi^2}{6} \times 24 }$ is related
to the world sheet zero point energy of 24 transversal oscillators.
$S_c$ indicates that the entropy  has been obtained by integrating only
over the common $S_2$ cut. If we take now the case of $d=2$: 
$\vec{Q}= (Q_1 , 0 , Q_2 , 0)$, $\vec{P}= (0 ,P_1 , 0 , P_2)$, 
$\vec{n}= (0, 0, 1 , 0)$, $\vec{p}= ( 0, 0 , 0 , 1)$ 
and all non-vanishing
charges large (to keep the 4d perturbation expansions under control),
the separate cases are: \vspace{3mm}

(i) $a=0$:  $S_c = 2 \pi \sqrt{Q_1 Q_2 P_1 P_2}$, \vspace{1mm}

(ii) $a=1$: $S_c = 4 \pi \sqrt{Q_1 Q_2 /2}$ ($P_1=P_2=0$, electric case),
\vspace{1mm}

(iii) $a=\sqrt{3}$: $S_c = 4 \pi \sqrt{Q_1}$ ($Q_2 = P_1 = P_2 =0$, electric
case) \vspace{2mm}

\noindent 
or identifying the charges (or harmonic functions) we get
\be150
S_c = 2 \pi \sqrt{1 +a^2} \, |Q|^{\frac{2}{1+a^2}} \ .
\ee
For the $a=1, \sqrt{3}$ cases we considered only the electric
part. Of course, via $S$-duality in 4 dimensions every electric
solution has its magnetic analogue with the charges $\vec{P} = - 
L \vec{Q}$. This solution gives the magnetic part of the entropy.
Note that the charge vectors are perpendicular to each other.
This is a result of the fact that all $U(1)$'s are related
to different directions in the internal space. In \cite{cv/ts},
in the case of $a=0$,
it has been argued that the other cases with arbitrary charge
vectors (and all charges large) correspond to a non-vanishing
axion in 4 dimensions. In this case the area formula has to
include a correction term ($\sim \vec{P} L \vec{Q}$). Since
the axion is related to a NS charge and we are considering only
RR intersections it is natural that we do not get this term.

In our approach the case $a=1/\sqrt{3}$ is special. The electric case
yields, integrating over $S_3$ an entropy density $S\sim \sqrt{Q^3}$
whereas the magnetic case leads, after integrating over $S_2$, to
$S\sim P^2$.  Recently, many authors have investigated the electric
case (see e.g.~\cite{st/va}). However, it does not fit into our
entropy formula (we have 3 intersecting branes in this case). After
restricting to the common $S_2$ cut we have $S\sim Q$ for the electric
case and $\sim P^2$ for the magnetic case. To get the right power of
charges, it seems that we have to take the average of the electric and
magnetic contributions. Note that the $a=1/\sqrt 3$ BH is also special
in the sense that it cannot be expressed by $D$-3-branes only. In
order to get a non-singular result we need to include the wave or
Taub-NUT soliton.  Thus, we can conclude that the formula (\ref{140})
describes the entropy for all BH's that can be expressed in a
non-singular way by $D$-3-branes only.

We have considered only the non-singular intersections yielding the
BH's in (\ref{020}).  Of course, there are other higher--dimensional
solutions yielding the same black holes after compactification to
$d=4$. But all these solutions remain singular or strongly
coupled near the horizon resulting in a vanishing entropy. Therefore,
from all possible states only the discussed intersections contribute
to the entropy counting. These non-singular states are preferred by
the system.
\bigskip

\noindent {\bf 4.\ Conclusions}
\bigskip

In this letter we have discussed extremal 4d Maxwell/scalar black
hole solutions with scalar coupling parameter $a$ 
near the horizon. Only the $a=0$ black hole has a
non-vanishing Bekenstein-Hawking entropy. All black hole solutions
that include a scalar field coupling have usually vanishing
entropy. This is a puzzle, since the entropy counts the states
and it should be possible to have different states like, e.g.~,
pure magnetic or pure electric configurations.
On the other side we know that all these solutions appear as
the compactification of higher--dimensional solutions. There are many
possibilities to describe these solutions as intersections of
$p$-branes in 10 dimensions. But for every black hole type there exist just
one possibility which is non-singular and for which one can define a
non-vanishing Bekenstein-Hawking entropy: (i) the 10d type IIB 3-brane
for $a=\sqrt{3}$, (ii) the 6d type IIB 1-brane for $a=1$ and (iii) 5d
RN solution for the $a=1/\sqrt{3}$ solution. The first 2 cases can be
understood as intersections of 3-branes only. The last case, however,
requires an additional wave or Taub-NUT soliton for the intersection.
Compactifying these objects over periodic arrays results in an
effective higher--dimensional solution near the horizon.
As a consequence, the singularities disappear \cite{gi/ho/to} yielding a
non-trivial entropy. The different cases have different spherical
symmetry near the horizon. Integrating over the horizon we have
given an entropy formula (\ref{140}) that covers the cases of
pure $D$-3-brane intersections ($a=0,1,\sqrt{3}$). This 
generalizes the entropy formula given in \cite{la/wi} to the case of
a single $D$-3-brane.
\bigskip

\noindent {\bf Acknowledgements}
\bigskip

The work of K.B.~is suported by the DFG. The work of E.B.~has been made
possible by a fellowship of the Royal Netherlands Academy of Arts and 
Sciences (KNAW).


\bigskip



\begin{thebibliography}{99}
\bibitem{ra}
J.~Rahmfeld, {\it Extremal black holes as bound states},
{\tt hep-th/9512089}.
\bibitem{cv/ts}
M.~Cvetic and A.A.~Tseytlin,
{\it
Solitonic strings and BPS saturated dyonic black holes},
{\tt hep-th/9512031}.
\bibitem{du/ra}
M.J.~Duff and J.~Rahmfeld, {\it Bound states of black holes and other
$P$-branes}, {\tt hep-th/9605085}.
\bibitem{gi/ho/to} G.W.~Gibbons, G.T.~Horowitz and P.K.~Townsend,
{\it Higher-dimensional resolution of dilaton black hole singularities},
{\tt hep-th/9410073}.
\bibitem{gi/gr/pe}
G.W.~Gibbons, M.B.~Green and M.J.~Perry,
{\it Instantons and seven-branes in type IIB superstring theory},
{\tt hep-th/9511080}.
\bibitem{pa/to} G.~Papadopoulos and P.K.~Townsend, 
{\it Intersecting $M$-branes}, {\tt hep-th/9603087}.
\bibitem{ts}
A. A. Tseytlin, {\it Harmonic superpositions of M-branes}, 
{\tt hep-th/9604035}; I.R.~Klebanov and A.A.~Tseytlin,
{\it Intersecting $M$--branes as four-dimensional black holes},
{\tt hep-th/9604166}.
\bibitem{ba/la}
V.~Balasubramanian and F.~Larsen, {\it On D-Branes and Black Holes in 
Four Dimensions}, {\tt hep-th/9604189}.
\bibitem{be/be/ja}
K.~Behrndt, E.~Bergshoeff and B.~Janssen,
{\it Intersecting $D$--Branes in Ten and Six Dimension},
{\tt hep-th/9604168}.
\bibitem{ga/ka/tr} 
J.P.~Gauntlett, D.A.~Kastor and J.~Traschen,
{\it Overlapping Branes in $M$--Theory}, {\tt hepth/9604179}.
\bibitem{kh/lu/po} N.~Khviengia, Z.~Khviengia, H.~L\"u and C.N.~Pope,
{\it Intersecting $M$--branes and Bound States}, {\tt hep-th/9605077}.
\bibitem{ga/ha/li}
J.P.~Gauntlett, J.A.~Harvey and J.T.~Liu,
{\it Magnetic monopoles in string theory},
{\tt hep-th/9211056}, B.~Harrington and H.~Shepard,
Phys.~Rev.~{\bf D17} (1978) 2122.
\bibitem{ho/se}
G.T.~Horowitz and A.~Sen, 
{\it Rotating black holes which saturate a Bogomol'nyi bound},
{\tt hep-th/9509108}.
\bibitem{ho/st}
G.T.~Horowitz and A.~Strominger, {\it Black strings and $p$-branes},
Nucl.Phys.~{\bf B360} (1991) 197.
\bibitem{fe/ka}
S.~Ferrara and R.~Kallosh,
{\it Supersymmetry and attractors}, {\tt hep-th/9602136};
{\it Universality of supersymmetric attractors}, {\tt hep-th/9603090}.
\bibitem{ho}
G.T.~Horowitz, 
{\it The origin of black hole entropy in string theory},
{\tt hep-th/9604051}.
\bibitem{gu/kl/pe}
S.S.~Gubser, I.R.~Klebanov and A.W.~Peet,
{\it Entropy and temperature of black 3-branes},
{\tt hep-th/9602135}.
\bibitem{la/wi}
F.~Larsen and F.~Wilczek, {\it Internal structure of black holes},
{\tt hep-th/9511064}
\bibitem{st/va}
A.~Strominger and C.~Vafa,
{\it Microscopic origin of the Bekenstein--Hawking entropy},
{\tt hep-th/9601029}.

\end{thebibliography}
\end{document}